**Comment on "Advanced field emission measurement techniques for research on modern cold cathode materials and their applications for transmission-type x-ray sources" [Rev. Sci. Instrum. 91, 083906 (2020)]**

Richard G. Forbes

AFFILIATION

Advanced Technology Institute and Department of Electrical and Electronic Engineering, University of Surrey, Guildford, Surrey, GU2 7XH, UK

r.forbes@trinity.cantab.net

ABSTRACT

This Comment suggests that technological field electron emission (FE) papers, such as the paper under discussion [P. Serbun et al.,, Rev. Sci. Instrum. **91**, 083906 (2020)], should use FE theory based on the 1956 work of Murphy and Good (MG), rather than a simplified version of FE theory based on the original 1928 work of Fowler and Nordheim (FN). Use of the 1928 theory is common practice in technological FE literature, but the MG treatment is known to be better physics than the FN treatment, which contains identifiable errors. The MG treatment predicts significantly higher emission current densities and currents for emitters than does the FN treatment. From the viewpoint of the research and development of electron sources, it is counterproductive (and unhelpful for non-experts) for the technological FE literature to use theory that undervalues the performance of field electron emitters.

Serbun et al. have recently provided a very useful review[1] concerning advanced field emission measurement techniques for research on modern cold cathode materials. In order to analyse their results, they make the usual assumption that their emitters can be modelled as if they were free-electron metals with a smooth planar surface of large lateral extent —so-called "smooth-planar-metal-like-emitter (SPME) methodology", and use an equation that is related to a simplified version[2] of an equation derived[3] by Fowler and Nordheim in 1928. The Serbun et al. equation, in the variant that relates current $I$ to voltage $V$, is given in the form

$$I = \frac{SA\beta^2 V^2}{\phi d^2} \exp\left[-\frac{B\phi^{3/2}d}{\beta V}\right], \tag{1}$$

where $\phi$ is the relevant local work function, and $A$ are $B$ the Fowler-Nordheim constants[4]. $V$ is the applied voltage between, and $d$ is the separation between, the anode and the emitter. $\beta$ is the effective field enhancement factor at the emitter tip, and $S$ is the formal emission area.

The slope $\Sigma$ of the corresponding Fowler-Nordheim (FN) plot is given by

$$\Sigma = \mathrm{d}[\ln\{I/V^2\}]/\mathrm{d}(1/V) = -\frac{B\phi^{3/2}d}{\beta}, \tag{2}$$

and the related extraction formula for the field enhancement factor is

$$\beta^{\mathrm{extr}} = -\frac{B\phi^{3/2}d}{\Sigma}. \tag{3}$$

If the parameters $V$ and $I$ are interpreted as the measured voltage and current, then the parameter $\beta$ is constant (and hence a useful characterization parameter) *only if* the measurement system is operating in so-called *ideal* fashion, where there are no "complications" such as those created by series resistance in the current path, current dependence in $\beta$, field dependence in emitter geometry due (for example) to Maxwell stress, or space-charge effects. Serbun et al. correctly apply eq. (1) only to those regions of Fowler-Nordheim plots where the emitters are apparently behaving (and very probably are behaving) in an ideal fashion.

Although use of eq. (1) for data-analysis purposes is relatively common in the field emission technological literature of the last 20 or so years, conceptual and mathematical errors in the thinking of Fowler and Nordheim[3] and in a related paper by Nordheim[5] were in fact discovered in the 1950s. An improved theory of field electron emission (FE) was developed by Murphy and Good[6] in 1956, based on the assumption that tunnelling takes place through a planar image-rounded tunnelling barrier, often now called a "Schottky-Nordheim (SN) barrier". Due to changes in conventions, their paper is now not easy to follow: Ref. 7 is a modern derivation.

The Murphy-Good (MG) FE equation equivalent to eq. (1) can usefully be written in the form

$$I = \frac{SA\beta^2 V^2}{\phi d^2} \exp\left[-v_F \frac{B\phi^{3/2} d}{\beta V}\right], \qquad (4)$$

where $v_F$ is the appropriate particular value (for a SN barrier defined by $\phi$ and the relevant local surface field $F$) of a well-defined[6] FE special mathematical function "v". (Here, the pre-factor $t^{-2}(y)$—typically of value around 0.9— that appears in the original version of MG theory has been assimilated into the formal emission area $S$.)

This equation generates a slightly modified extraction formula for the field enhancement factor $\beta$, namely

$$\beta^{\text{extr}} = -\frac{s_t B \phi^{3/2} d}{\Sigma}, \qquad (5)$$

where $s_t$ is the appropriate value of a slope correction function first calculated correctly[8] in 1953.

For an emitter with $\phi$=4.50 eV, the value of $v_F$ typically lies in or near the range 0.6 to 0.7, and $s_t$ can be adequately approximated as 0.95. For a given assumed value of $S$, eq. (4) predicts current values that are significantly higher[9] than those predicted by eq. (1), by a factor typically of order 300.

As already indicated, at present it is often community practice in technological field emission to state an equation based on the original 1928 treatment, rather than an equation based on the corrected treatment published in 1956. However, a recent careful re-examination[9] of the physics underlying the two treatments has confirmed that the 1956 treatment is in fact better physics, as long assumed by most theoretical experts in FE. It therefore seems that a change in practice, to a situation where some appropriate version of Murphy-Good FE theory is stated in technological FE papers, is strongly indicated. The Serbun et al. paper has been published in a "Review" journal, so it seems particularly relevant to make this point in connection with it.

It needs stressing that the 5% change in extracted value of field enhancement factor is not important in the absolute scale of things. Rather, the main problem is the repeated literature statement of an equation that, as compared with better physics available within the framework of SPME methodology, seriously under-predicts the performance of a field emitter of given assumed emission area. Also, it can be shown[10] that, if eq. (4) were applied to an ideal FN plot in order to extract an "experimental" value of emission area, then the extracted area value would be very different (by a factor that could be around 100) from that extracted by using eq. (1).

More generally, there seems no good scientific reason to continue to use 90-year-old theory, when a theory based on apparently better physics has been available for more than 60 years (and when there is now additional support for thinking that it is better physics), and when the better physics predicts significantly higher emitter performance. From the viewpoint of the research and development of electron sources, it seems counterproductive to state an equation that significantly under-predicts and under-values the performance of field electron emitters. In particular, this practice is misleading for non-experts.

It is also true that it is long overdue that SPME methodology should be superseded by improved data-analysis methodologies that take the actual shapes of point-like field emitters and the existence of atomic structure into detailed account. However, improved methodologies of this kind are not yet well established. In the author's view, it is likely to be easier to move on from SPME methodology if technological FE literature first moves to the situation where all papers use essentially the same version of SPME methodology, based on Murphy-Good FE theory.